# Synergistic effect of oxygen and water on the environmental reactivity of 2D layered GeAs

*Luca Persichetti[1,±], Giacomo Giorgi[2,3,4,5], Luca Lozzi[6], Maurizio Passacantando[6,7], Fabrice Bournel[8,9], Jean-Jacques Gallet[8,9], Luca Camilli[1,*]*

[1]*Dipartimento di Fisica, Università di Roma "Tor Vergata", Via Della Ricerca Scientifica, 1- 00133 Rome, Italy*

[2]*Department of Civil and Environmental Engineering (DICA), University of Perugia, Via G. Duranti 93, 06125, Perugia, Italy;*

[3]*CIRIAF – Interuniversity Research Centre University of Perugia Via G. Duranti 93, 06125 Perugia, Italy*

[4]*CNR-SCITEC, 06123 Perugia, Italy*

[5]*Centro S3, CNR-Istituto Nanoscienze, Via G. Campi 213/a, Modena, 41125, Italy.*

[6]*Department of Physical and Chemical Science, University of L'Aquila, via Vetoio, Coppito, 67100 L'Aquila, Italy*

[7]*CNR-SPIN L'Aquila, via Vetoio, Coppito 67100, L'Aquila, Italy*

[8]*Sorbonne Université, CNRS, Laboratoire de Chimie Physique-Matière et Rayonnement, Campus Curie, UMR 7614, 4 place Jussieu, 75005 Paris, France*

[9]*Synchrotron SOLEIL, L'orme des Merisiers, B.P. 48, Saint Aubin, Gif-sur-Yvette Cedex 91192, France*

We investigated the reactivity of layered GeAs in the presence of oxygen and/or water using synchrotron-based X-ray photoelectron spectroscopy and *ab initio* calculations. By performing experiments at near-ambient pressure (up to 20 mbar), we gained detailed insights into the material's stability under realistic conditions. GeAs showed limited reactivity with dry $O_2$ and de-aerated $H_2O$. However, a small amount of humidity ($R_w = 0.5\%$ at $T = 20°C$) in an $O_2$ atmosphere significantly enhanced reactivity. This synergistic effect was well captured by density functional theory calculations, which revealed a strongly exothermic formation energy for the simultaneous chemisorption of $O_2$ and $H_2O$, compared to the adsorption of each molecule individually.

[*]corresponding authors:

[±]luca.persichetti@uniroma2.it (Luca Persichetti)

[*]luca.camilli@uniroma2.eu (Luca Camilli)

# 1. Introduction

The world of two-dimensional (2D) materials has become a rich playground for innovation, offering a unique combination of properties[1-3]. From electronics to catalysis and energy storage, 2D materials hold promise for revolutionizing various technological fields[4-6]. However, their journey to practical applications is not without challenges. The exceptional properties of 2D materials are ofttimes highly susceptible to the environment, particularly when it comes to oxidation and water interactions[7-13]. Thence, investigating the reactivity of 2D materials with air gases—such as oxygen, oxygen in the presence of moisture, or water—is crucial for understanding the chemistry behind material degradation and for developing effective protection strategies to enable their use in ambient conditions.

As a model system for investigating the environmental stability, we will focus on germanium arsenide (GeAs). GeAs is a p-type layered semiconductor with a sizeable indirect band gap reaching about 2.1 eV for the monolayer (ML)[14, 15], with the hole carrier mobility approaching 100 $cm^2/(V \cdot S)$[16] and excellent thermoelectric properties[17]. It belongs to the group IV-V pnictides family and crystallizes into a monoclinic C2/m layered structure[18] with each layer terminated by arsenic atoms and an interlayer distance of 660 pm[16, 19] (Fig. 1). The structure is strongly anisotropic with two types of Ge-Ge bonds being either parallel or perpendicular to the layer plane[20, 21], giving rise to pronounced anisotropy in the in-plane electronic and optical properties, such as polarization-sensitive photoresponse and anisotropic mobility[19, 22-25]. Field-effect transistors (FETs) and photodetectors based on few-layer GeAs have been showing promising performances[16, 26-30].

However, there is some controversy in literature when it comes to the stability of GeAs devices in environmental conditions. Indeed, while some reports describe GeAs as sufficiently stable [27, 31], others observe that the topmost layer undergoes significant degradation when exposed to air[32], suggesting that the material's intrinsic properties may be altered in devices operating under ambient conditions. Because of the absence of specific studies focusing on understanding the chemical reactivity of GeAs in realistic environmental conditions, it was unclear until now which specific air component contributed to the material's degradation. In this paper, we address this knowledge gap by exploring the oxidation of GeAs induced by dry $O_2$, $H_2O$ vapor, or a combination of the two in a humid $O_2$ atmosphere. Notably, we employed synchrotron-based X-ray photoelectron spectroscopy (XPS) performed in near-to-ambient pressure (NAP) conditions and density functional theory (DFT) calculations to elucidate the oxidation mechanism of GeAs. Our findings emphasized the crucial synergistic effect of oxygen and water in enhancing the reactivity of GeAs, leading to significantly higher oxidation when both reactants were present simultaneously when compared to the oxidation with the individual reactants. This knowledge about the interaction of 2D layered semiconductors with environmental gases is essential for devices operating under realistic conditions.

## 2. Materials and methods

Experiments were performed with the NAP-XPS endstation of Sorbonne Université at the TEMPO beamline (SOLEIL synchrotron facility, St-Aubin, France) on a single crystal GeAs (2D semiconductors) which was cleaved in the load-lock chamber of the ultra-high-vacuum (UHV) system at a pressure $<5\times10^{-8}$ Torr and immediately transferred into the analysis chamber at a base pressure $\sim 1\times10^{-9}$ Torr. NAP-XPS was carried out at room temperature using a hemispherical electron analyzer (Phoibos 150 NAP, SPECS) with a cone aperture of 300 µm positioned approximately 1 mm from the sample surface. The analyzer axis was perpendicular to the sample surface while the windowless beam entrance direction formed a 54° angle with respect to the analyzer axis. A differential pumping stage separated the back-filled analysis chamber from the analyzer and the beamline. The escaped photoelectrons were focused using an electrostatic lens system.

Gases were dosed into the analysis chamber using dedicated gas lines and leak valves, depending on the specific experiment, either with the sample under X-ray irradiation (light-on condition) or with the beam blanked during exposure (dark condition). We investigated the oxidation of GeAs kept at room temperature in a dry atmosphere of pure $O_2$, as well as in a wet atmosphere of $O_2/H_2O$ in the ratio 12:1. The gases were dosed at a pressure ranging between 0.1-17 Torr, as described in the Results section, and monitored by a quadrupole mass spectrometer (QMS) in the second differential pumping stage of the electron analyzer. Exposure dosage is given in langmuir (1 L= $10^{-6}$ Torr×s). Dry oxidation was studied before water introduction in the analysis chamber to avoid cross-contamination. Water vapor was introduced through a leak valve from a liquid reservoir containing ultrapure water (>18 MΩ×cm), which was carefully degassed to obtain deaerated water by multiple freeze–pump–thaw cycles. The relative humidity is defined as $R_w = 100(p_w/p_0)$, where $p_w$ is the water vapor pressure in the analysis chamber and $p_0$ the equilibrium water vapor pressure at the sample temperature $T$. At $T= 20$ °C, $p_0$ is 17. 5 Torr[33].

We collected the Ge 3d and As 3d core levels within a single spectrum sweep, so that the two are always normalized to each other. Unless otherwise stated, to monitor the formation of ultra-thin oxide layers, we used a photon energy $E_p= 350$ eV for maximizing surface sensitivity. Spectra acquired at higher photon energies were used for evaluating changes along the thickness of the GeAs sample. All spectra were energy-calibrated with respect to the position of Au $4f_{7/2}$ photoelectron corresponding to a binding energy (BE) of 84.0 eV[34]. Peak fitting was performed using the KolXPD software by using Voigt functions after a Shirley type background subtraction. The following constraints were used: (*i*) same spin-orbit splitting and same branching ratio for each element-specific (Ge or As) component; (*ii*) same Gaussian width for all the components depending on the analysis chamber pressure during acquisition; (*iii*) Lorentzian width and energy position were fixed for each Ge (As) component in all the experiments/conditions. In particular, the spin orbit splitting of the characteristic $3d_{5/2}$ and $3d_{3/2}$ doublet was set to 0.58 eV for Ge[35], and to 0.68 eV for As[36].

*Theoretical Calculations*

DFT calculations were performed using the VASP code[37-40]. The Generalized Gradient Approximation (GGA) of Perdew, Burke, and Ernzerhof (PBE) was exploited to solve the exchange-correlation part of the potential[41]. The projector-augmented wave (PAW) potentials[42], with an energy cutoff of 500 eV, was adopted. The D3(BJ) dispersion correction contribution was included in the calculations[43]. Geometries were considered converged when forces were lower than 0.02 eV/Å.

Because we were interested in the first attacks of the oxidant species on the GeAs surface, we considered one ML of GeAs with a 1x4 supercell (see Fig. 1), adding a relevant amount of vacuum along the non-periodic direction to avoid any possible spurious interactions between replicas. The optimized supercell lattice lateral parameters are $a$= 21.89 Å, $b$=15.27 Å.

The formation energy, $E_f$, of the product of the oxidation reaction was evaluated following the general equation:

$$E_f = E_{GeAs\_X} - (E_{GeAs} + E_X), \quad (1)$$

where $E_{GeAs\_X}$ is the energy of the oxidation product, $E_{GeAs}$ is the energy of the GeAs ML, and $E_X$ is the energy of the oxidant species (X=$O_2$; $H_2O$).

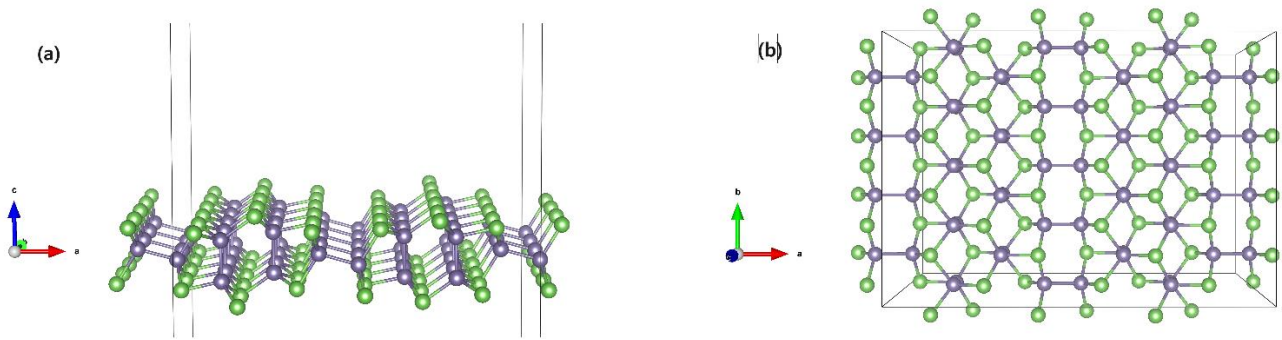

**Fig 1.** (a) Lateral and (b) top view of the PAW/PBE optimized (1×4) supercell of the GeAs ML (green: As; gray: Ge atoms).

## 3. Results

*$O_2$ oxidation of GeAs*

In this section, we present XPS data for the oxidation of GeAs in dry $O_2$ and humid ($O_2/H_2O$) atmosphere. The effect of soft X-ray irradiation on $O_2$ oxidation during *in operando* NAP-XPS measurements is also evaluated. The results are summarized in Fig. 2 where the Ge 3d and As 3d core levels of pristine GeAs are compared with those after exposure. As reported in Figs. 2(a) and 2(b), pristine GeAs shows a single component ($Ge^0$ and $As^0$, respectively) for both the Ge 3d and As 3d, the energy position of which is 29.9 eV for $Ge^0$ and 41.3 eV for $As^0$, in agreement with previous reports[44, 45]. The ratio between the total area of the As 3d and Ge 3d peaks ($A_{As}^{tot}/A_{Ge}^{tot}$= 1.2) closely matches that of the corresponding photoionization cross-sections (i.e. 1.19), confirming the 1:1 stoichiometry of pristine GeAs.

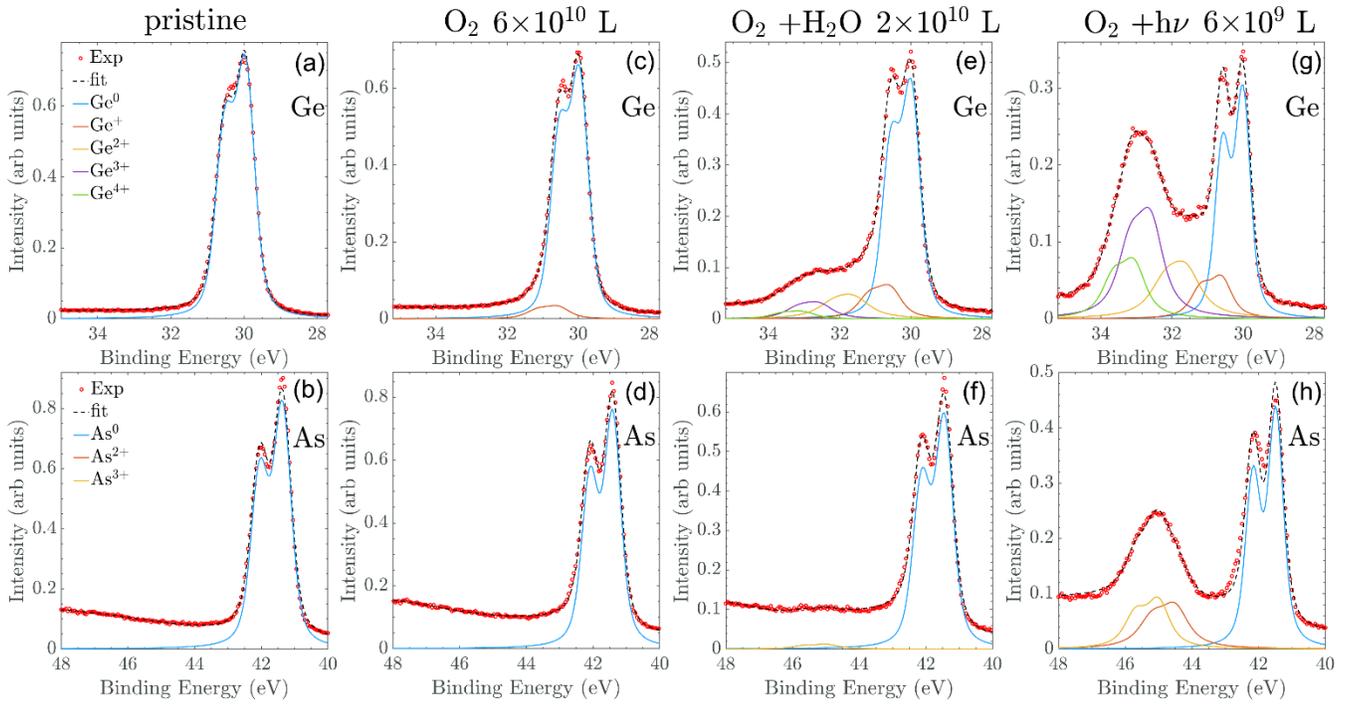

**Fig. 2.** XPS spectra of (first row) Ge 3d and (second row) As 3d core levels for: (a, b) pristine GeAs film; (c, d) GeAs after exposure to $6.0\times10^{10}$ L of $O_2$; (e, f) GeAs after exposure to $2.0\times10^{10}$ L of $O_2+H_2O$ ($R_w$=0.5%); (g, h) GeAs after exposure to $6.0\times10^{9}$ L of $O_2$ under X-ray irradiation. Spectra are measured with $E_p$= 350 eV and normalized to the total Ge 3d area of each sample.

As evident in Figs. 2(c) to 2(h), the degree of oxidation of GeAs strongly depends on the conditions of the $O_2$ exposure. In dry $O_2$ atmosphere [Figs. 2(c) and 2(d)], GeAs is almost not oxidized even at a massive dosage of $6\times10^{10}$ L, obtained with 1 hour exposure at 17 Torr. Indeed, no components associated to arsenic oxides are present, while only a minor component for the $Ge^+$ oxidation state is found at a binding energy shifted by $\Delta$=+0.6 eV[44, 46] with respect to $Ge^0$. The stoichiometry obtained from the ratio $A_{As^{tot}}/A_{Ge^{tot}}$ remains 1:1 as for pristine GeAs, while the attenuation of the unoxidized $Ge^0$ component ($A_{Ge^0}/A_{Ge^{tot}}$) decreases to 0.93, which, considering the effective inelastic mean free path $\lambda$= 0.56 nm for 320 eV kinetic energy photoelectrons, would give an equivalent thickness $L$= 40 pm for a uniform oxide adlayer. This, together with the fact that $Ge^+$ suboxide is typically the dominant one at the very surface of $GeO_x$[35, 47], indicates that the oxidation in dry $O_2$ remains limited and extremely superficial. This is confirmed by data obtained with $E_p$= 500 eV (see Fig. S1 in supporting information), from which the ratio $A_{Ge^0}/A_{Ge^{tot}}$ is even higher, reaching 0.95 ($L$= 37 pm). As observed previously with 350 eV photon energy, the Ge 3d spectrum at $E_p$= 500 eV shows only Ge$^+$, with no detectable AsO$_x$ components.

Because it has been observed that the first monolayer of GeAs strongly degrades in air[32, 45], it is interesting to evaluate the reaction with humid O$_2$. Figures 2(e) and 2(f) show Ge 3d and As 3d XPS spectra after exposing pristine GeAs to $2\times10^{10}$ L of O$_2$/H$_2$O in the ratio 12:1. This is achieved by dosing water vapor at a controlled partial pressure of 90 mTorr in a O$_2$ partial pressure of 1.13 Torr ($R_w$= 0.5% at T= 20 °C) for about 5 hours. Despite the low relative humidity, the oxidation of GeAs in humid O$_2$ atmosphere is dramatically stronger than that observed for dry O$_2$, even at a 1/3 of the dosage. Indeed, the attenuation of the unoxidized Ge$^0$ component is now 0.54, giving an equivalent Ge oxide thickness $L$= 350 pm, almost an order of magnitude higher than what was found with dry O$_2$. Yet, it is worth noting that this value, i.e. 350 pm, is still smaller than the nominal GeAs monolayer thickness of 480 pm. In contrast to dry O$_2$, we observe all the oxidation states associated to GeO$_x$. The dominant sub-oxide components are Ge$^+$ (Ge$_2$O) and Ge$^{2+}$ (GeO, $\Delta$= +1.6 eV)[35], the area of each being about 20% of that of Ge$^0$, while the area of the Ge$^{3+}$ (Ge$_2$O$_3$, $\Delta$= +2.5 eV)[35] is about 10% lower. It is worth noting that germanium also reaches the Ge$^{4+}$ oxidation state ($\Delta$= +3.1 eV)[35], characteristic of bulk GeO$_2$. Similar results in terms of relative abundance of the GeO$_x$ components are obtained for $E_p$= 500 eV (Fig. S2 in supplementary information). By looking at the As 3d core level [Fig. 2(f)], a small component (~3% of the As$^0$ area) is observed at larger binding energy with respect to the As$^0$ peak. The energy shift ($\Delta$= +3.52 eV) between the two is compatible with As$^{3+}$ in As$_2$O$_3$[48]. It is interesting to note that while the total germanium area $A_{Ge^{tot}}$ remains constant (within 4%) before and after the exposure, the total arsenic area $A_{As^{tot}}$ is reduced, dropping to 77% of its initial value. As already observed for dry O$_2$ under illumination[44], this means that while germanium is oxidized forming an oxide on the GeAs surface, most of the superficial arsenic involved in the reaction is lost. Given an escape depth of approximately 3$\lambda$ (about 1.68 nm), corresponding to roughly 2.5 monolayers of GeAs, it is estimated that about 5/6 of the arsenic atoms on the top surface of the first layer (Fig. 1) form volatile $AsO_x$ species, which sublimate in a vacuum[49]. Overall, it is evident that the presence of a small amount of water significantly enhances the reactivity of GeAs with $O_2$.

Lastly, we examine the effects of dry $O_2$ exposure under X-ray illumination by continuously acquiring XPS spectra within the energy range of the Ge 3d and As 3d core levels in an $O_2$ atmosphere. Previous results obtained under an $O_2$ partial pressure of $1.3\times10^{-6}$ Torr for dosage up to $10^4$ L showed that oxidation was dramatically accelerated by the photon beam[44]. To go beyond this range, here we exposed pristine GeAs for 2 hours at 0.83 Torr of $O_2$ reaching a dosage of $6\times10^9$ L [Figs. 2(g), 2(h)]. As evident from the comparison with Figs. 2(c, d), although the dosage with $O_2$ under the beam is about 10 times lower than that reached with dry $O_2$ in dark conditions, both Ge and As are dramatically more oxidized under illumination. Notably, the attenuation of the unoxidized $Ge^0$ component after oxidation under the beam is 0.35, resulting in an equivalent oxide layer thickness $L$= 590 pm. Compared to the oxidation experiments in dark condition (without illumination), this value is almost 15 times higher than that for dry $O_2$ and 1.7 times higher than that for humid $O_2$. However, it is important to point out that the oxide thickness still does not exceed the first monolayer thickness. In the oxidation experiment under beam, at $6\times10^9$ L, $Ge^{3+}$ is the dominant $GeO_x$ component, with its area being 75% of that of $Ge^0$. The area ratio of the $Ge^{2+}$ ($Ge^+$) suboxide component to $Ge^0$ is 0.45 (0.25). For $AsO_x$, we observe both the $As^{3+}$ ($As_2O_3$) component and an additional one at lower binding energy, shifted with respect to the $As^0$ peak of $\Delta$= 3.0 eV, which we attributed to the $As^{2+}$ oxidation state (AsO)[50]. Also in this case, the oxidation reaction is accompanied by a loss of arsenic. The full dataset of XPS spectra acquired during exposure to $O_2$ is displayed in Figs. 3(a) and 3(b).

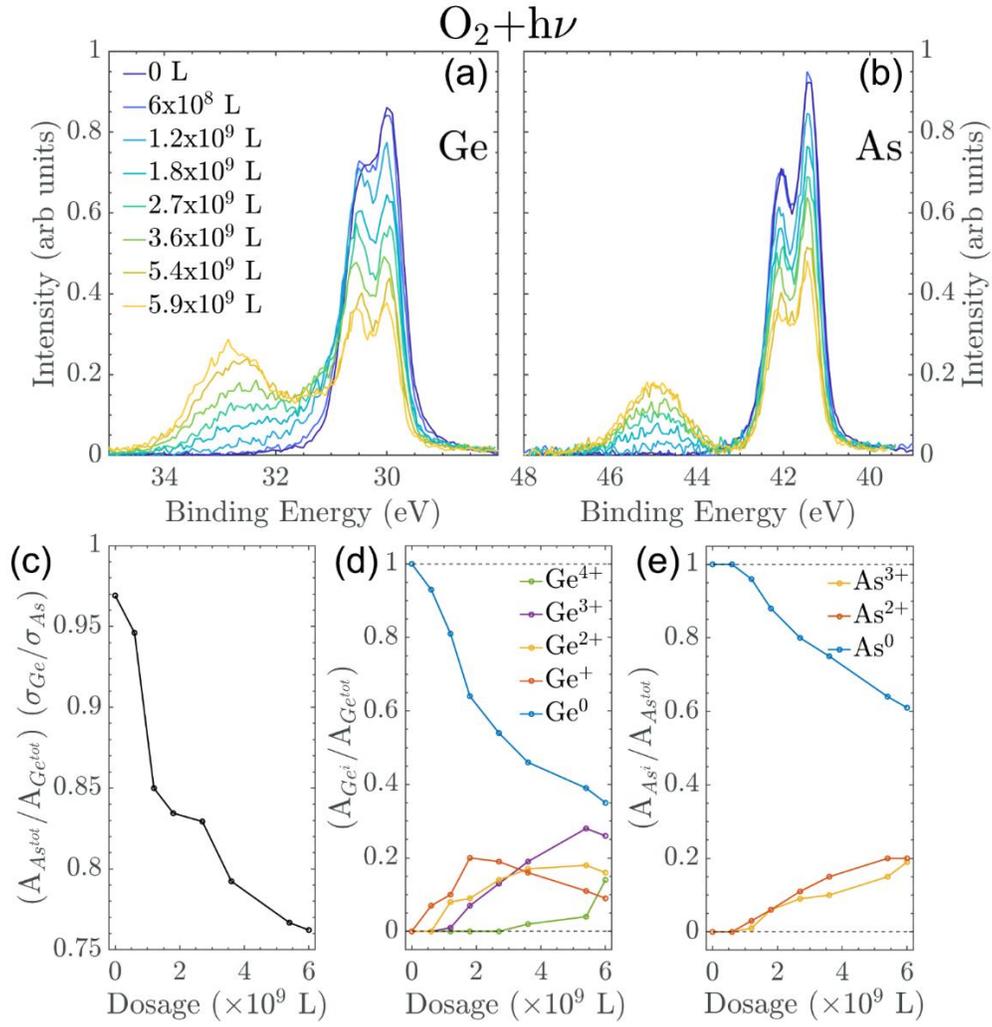

**Fig. 3. (a, b) NAP-XPS spectra acquired while exposing GeAs to 0.83 Torr of O$_2$. $E_p$= 350 eV. Spectra are normalized to the total Ge 3d area. (a) Ge 3d and (b) As 3d core levels at increasing dosage. (c-e) Evolution with dosage of: (c) stoichiometry obtained from the ratio of the areas of the As 3d and Ge 3d peaks normalized by the corresponding photoionization cross sections; (d) area of each Ge 3d fitting component with respect to the total Ge 3d area; (e) area of each As 3d fitting component with respect to the total As 3d area.**

In particular, we evaluated, as a function of the dosage, the ratio of the total areas of the As 3d and Ge 3d peaks normalized by the corresponding photoionization cross-sections [Fig. 3(c)]. The ratio defines the relative amount of arsenic to germanium present in the sample volume within the escape depth. Its value drops by almost 22% at 6×10$^9$ L with respect to that in the sample before oxidation. Figure 3(d) [3(e)] shows the evolution of the area of each component of the Ge 3d [As 3d] XPS peak, normalized to $A_{Ge^{tot}}$ [$A_{As^{tot}}$], as a function of dosage during the oxidation under X-ray irradiation. At the onset of the oxidation experiment, the dominant suboxide component is Ge$^+$, while, at a later stage, Ge$^{2+}$ and Ge$^{3+}$ become the most abundant. The $Ge^{4+}$ component (typical of bulk $GeO_2$) only appears at high dosage ($2.7\times10^9$ L). Regarding the arsenic components, it is noteworthy that arsenic oxide ($As^{2+}$ and, to a lesser extent, $As^{3+}$) starts to appear at a dosage of $1.2\times10^9$ L. At this point, as shown in Fig. 3(c), the loss of arsenic atoms is around 16%, indicating the removal of slightly less than 5/6 of the topmost arsenic atoms in the first layer (considering the total number of arsenic atoms within the escape depth). At the same dosage, the attenuation of the unoxidized $Ge^0$ component is 0.81 [Fig. 3(d)], giving an equivalent thickness of the germanium oxide of about 110 pm for a uniform layer. This value corresponds to about 1/5 of the GeAs monolayer thickness.

It can thus be inferred that, upon reaction with $O_2$, a significant portion of the arsenic atoms is initially displaced from the uppermost As layer of the topmost GeAs layer. Subsequently, oxygen starts bonding with the underlying germanium atoms to form a layer of germanium oxide. The formation of non-volatile $GeO_x$ impedes the loss of further arsenic from the bottom As layer in the first GeAs layer, and, thus, $AsO_x$ is finally formed. This could explain why $AsO_x$ is formed much later in the reaction with respect to $GeO_x$. Note also that at the largest dosage of $6\times10^9$ L, the loss of arsenic corresponds to the entire uppermost arsenic layer in the topmost GeAs layer. The same findings are obtained for $E_p=700$ eV where the electron escape depth reaches a thickness of 4 GeAs layers, confirming that only the surface-most arsenic atoms are lost. Similar results were also observed in another experiment where the oxidation under the beam was performed in a lower $O_2$ partial pressure of 0.15 Torr, up to a dosage of $1\times10^9$ L (see Fig. S3 in supplementary information).

Note that the As 3d and Ge 3d XPS core level spectra acquired on a sample location not being irradiated during the $O_2$ exposure were almost unchanged compared to those measured before the exposure (see Fig. S4 in supplementary information), confirming that GeAs is essentially not reacting in a dry $O_2$ atmosphere.

*Reaction of GeAs with $H_2O$*

We investigated the reaction of pristine GeAs with pure water by dosing, in dark conditions, $H_2O$ vapor at 15 Torr ($R_w$= 85% at 20 °C). In Figs. 4 and 5, we report the Ge 3d and As 3d core level XPS spectra obtained with $E_p$= 350 eV and 500 eV, respectively. At a dosage of $5.2\times10^{10}$ L, similar to that used for $O_2$ oxidation [Figures 2(c) and 2(d)], the attenuation of the $Ge^0$ unoxidized component is 0.90 (0.98) for $H_2O$ at $E_p$= 350 eV (500 eV), while it was 0.93 (0.95) for $O_2$. Therefore, both water and oxygen exposures result in similar, low attenuation of the $Ge^0$ unoxidized component in GeAs. This indicates that the equivalent germanium oxide thickness is shallow in both cases, with values less than 40 pm (averaged across the two photon energies).

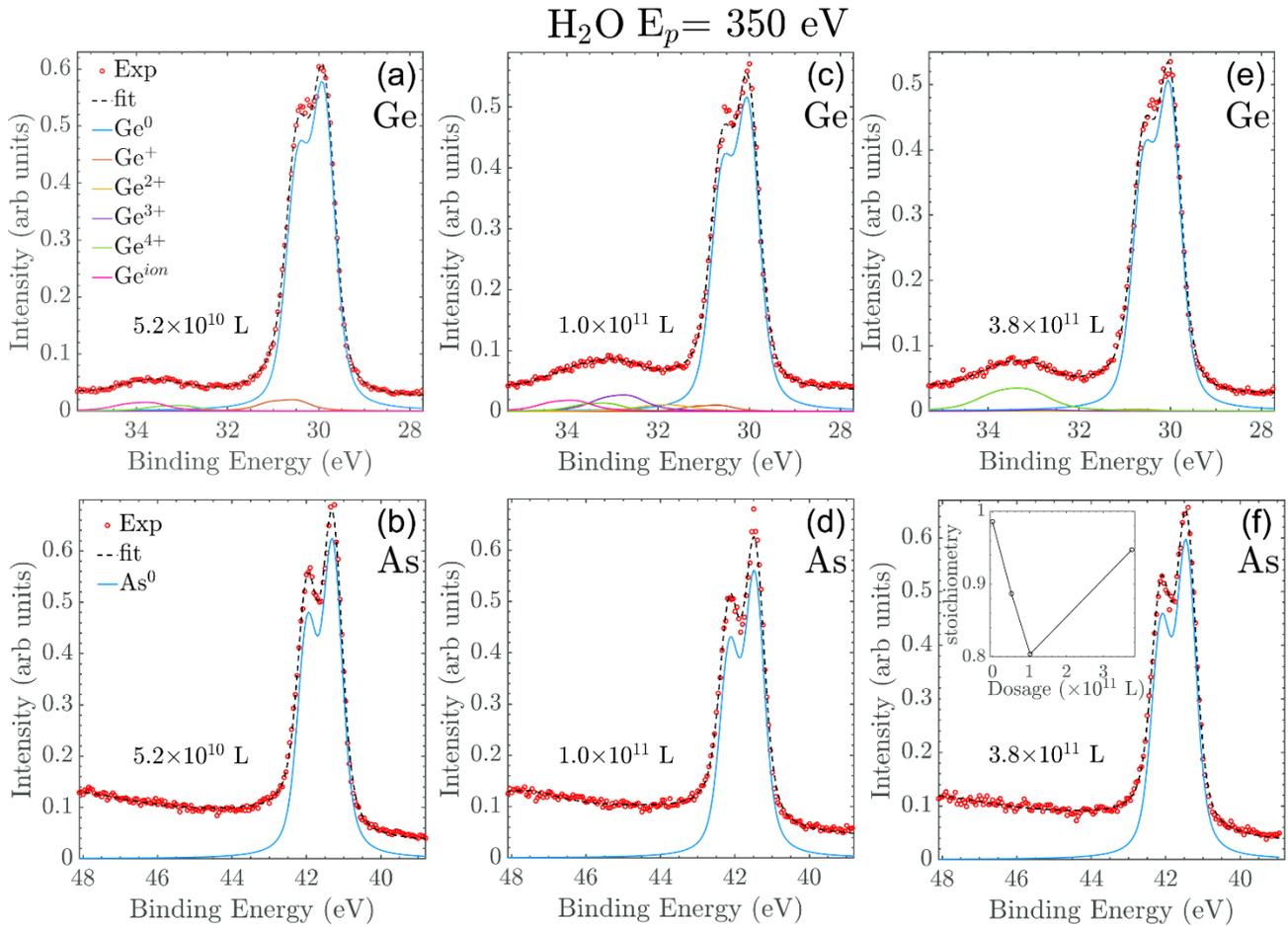

**Fig. 4.** XPS spectra of (top row) Ge 3d and (bottom row) As 3d core levels of GeAs after exposure to $H_2O$ at increasing dosage: (a, b) $5.2\times10^{10}$ L; (c, d) $1.0\times10^{11}$ L; (e, f) $3.8\times10^{11}$ L. Spectra are measured with $E_p$= 350 eV and normalized to the total Ge 3d area at each dosage. In the inset of panel (f), we display the evolution with dosage of the stoichiometry obtained from the ratio of the areas of the As 3d and Ge 3d peaks normalized by the corresponding photoionization cross sections.

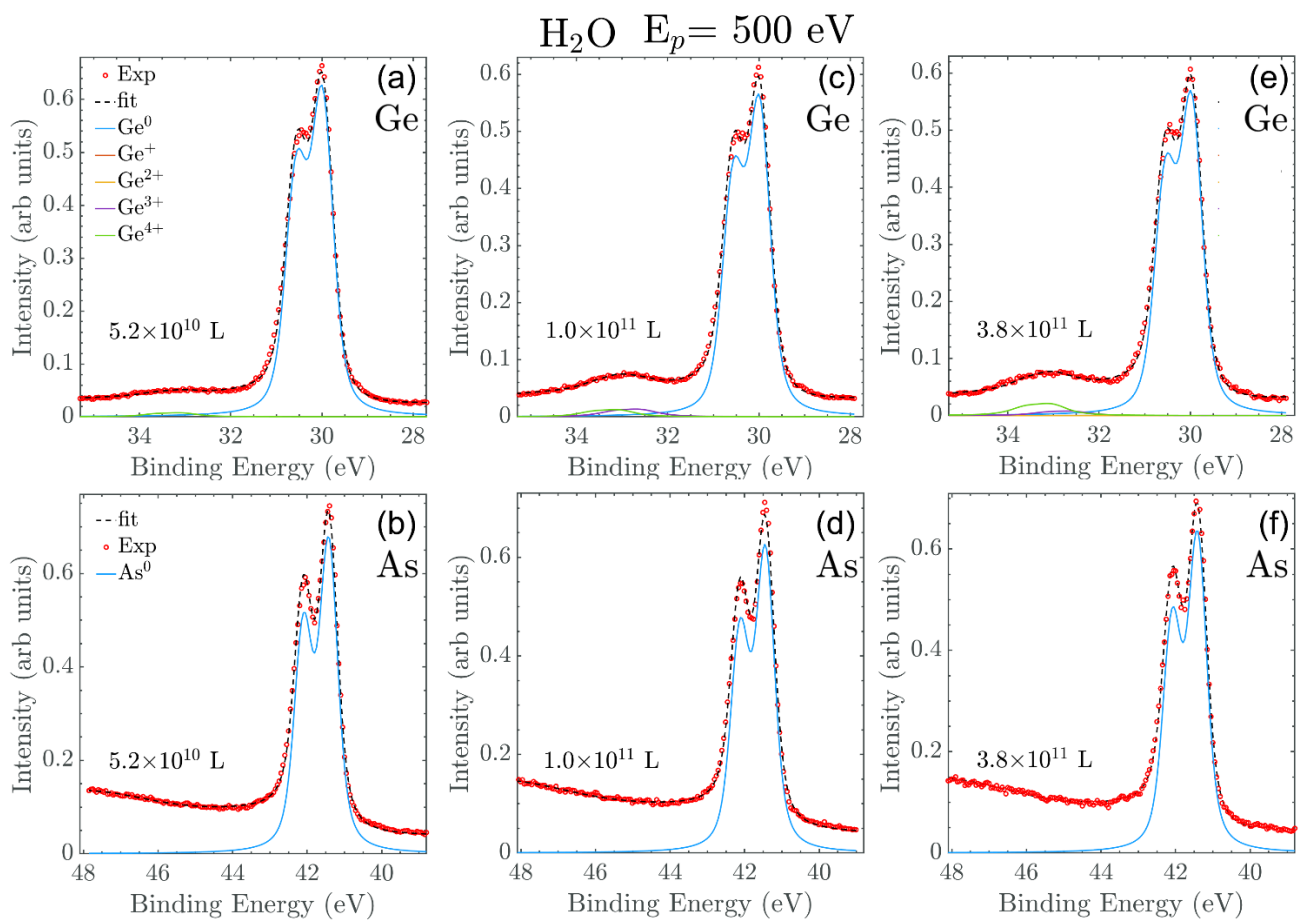

**Fig. 5.** XPS spectra of (top row) Ge 3d and (bottom row) As 3d core levels of GeAs after exposure to $H_2O$ at increasing dosage: (a, b) $5.2\times10^{10}$ L; (c, d) $1.0\times10^{11}$ L; (e, f) $3.8\times10^{11}$ L. Spectra are measured with $E_p$= 500 eV and normalized to the total Ge 3d area at each dosage.

Interestingly, when exposing pristine GeAs to $H_2O$, we found different spectroscopic fingerprints in the XPS data depending on the photon energy used, in contrast to what observed previously for $O_2$ oxidation. This is evident by comparing Figs. 4 and 5. At the lowest dosage of $5.2\times10^{10}$ L, the XPS Ge 3d core level spectrum with $E_p$= 500 eV shows almost no oxide components after water exposure [Fig. 5(a)], while some distinctive features emerge with $E_p$= 350 eV [Fig. 4(a)]. The dominant XPS components are in this case $Ge^+$ and a component (named as $Ge^{ion}$) with a larger binding energy than $Ge^{4+}$, showing $\Delta$= +3.75 eV. Its binding energy could be explained with the presence of ionic surface complexes, such as $Ge(OH)_2^{2+}$ or $-Ge-Ge^+-(OH)_2$ species. Their formation has been observed for germanium exposed to liquid water[51]. In our conditions, the vapor pressure of water is 15 Torr, compared to the saturation vapor pressure ranging between 12.8 and 17.5 Torr in the temperature interval 15- 20 °C[33]. Therefore, the formation of liquid water on top of our sample appears reasonable.

For larger dosages, at $1.0\times10^{11}$ L, the dominant component observed with $E_p$= 350 eV is the $Ge^{3+}$, whilst it becomes the $Ge^{4+}$ one at $3.8\times10^{11}$ L [Figs. 4(c) and 4(e)]. At the same time, the $Ge^{ion}$ component progressively disappears. We can therefore hint that, as the dosage increases, the $Ge^{ion}$ surface component is converted into $Ge^{3+}$ and later into $Ge^{4+}$. At the same dosages, for $E_p$= 500 eV, only $Ge^{3+}$ and $Ge^{4+}$ are present while no $Ge^{ion}$ component is observed [Figs. 5(c) and 5(e)], confirming that the latter is only present at the very surface of the GeAs sample. It is also interesting to observe that, with $E_p$= 350 eV, the attenuation of the unoxidized $Ge^0$ component changes from 0.90 ($L$= 59 pm) to 0.82 ($L$= 110 pm) when the dosage is increased from $5.2\times10^{10}$ L to $1.0\times10^{11}$ L, but then it drops again to 0.88 ($L$= 72 pm) at the largest studied dosage of $3.8\times10^{11}$ L. Here, the $Ge^{4+}$ component is the only one observed at both photon energies. This finding suggests that the reduction in equivalent germanium oxide thickness is due to the removal of $GeO_2$, which is water soluble[45, 51, 52], starting at the surface. By looking at the changes of the As 3d core level XPS spectra as a function of water vapor dosage, we observe that arsenic oxide is not observed [see Figs. 4(b), 4(d), 4(f) and Figs. 5(b), 5(d), 5(f)]. Previously, we noticed that $AsO_x$ started forming only when the germanium oxide thickness reached about 1/5 of a single GeAs layer. In the case of oxidation with water, this value (about 110 pm) is reached at $1.0\times10^{11}$ L. However, at larger dosage, the germanium oxide thickness again decreases due to the dissolution of water-soluble $Ge^{4+}$ oxide species. The absence of a germanium oxide layer being thick enough to impede the loss of volatile $AsO_x$ could explain why no arsenic oxide is observed, even at the largest dosage of $3.8\times10^{11}$ L, as opposite to the case of $O_2$ oxidation. By using the most surface-sensitive photon energy (350 eV), it is interesting to evaluate the ratio of the total areas of the As 3d and Ge 3d components normalized by the corresponding photoionization cross-sections [displayed as an inset in Fig. 4(f)]. As previously observed, the value (0.99) of the ratio before the exposure closely matches the GeAs stoichiometry. The ratio is reduced to 0.80 at a dosage of $1.0\times10^{11}$ L. Taking into account the escape depth, this value corresponds to the loss of most of the top arsenic atoms in the first monolayer. At the largest dosage, the As-to-Ge stoichiometry ratio rises again to about 0.95. This can be explained considering the removal of water soluble $Ge^{4+}$ species observed at the same dosage and it is therefore a confirmation of the attack of these species by water[45].

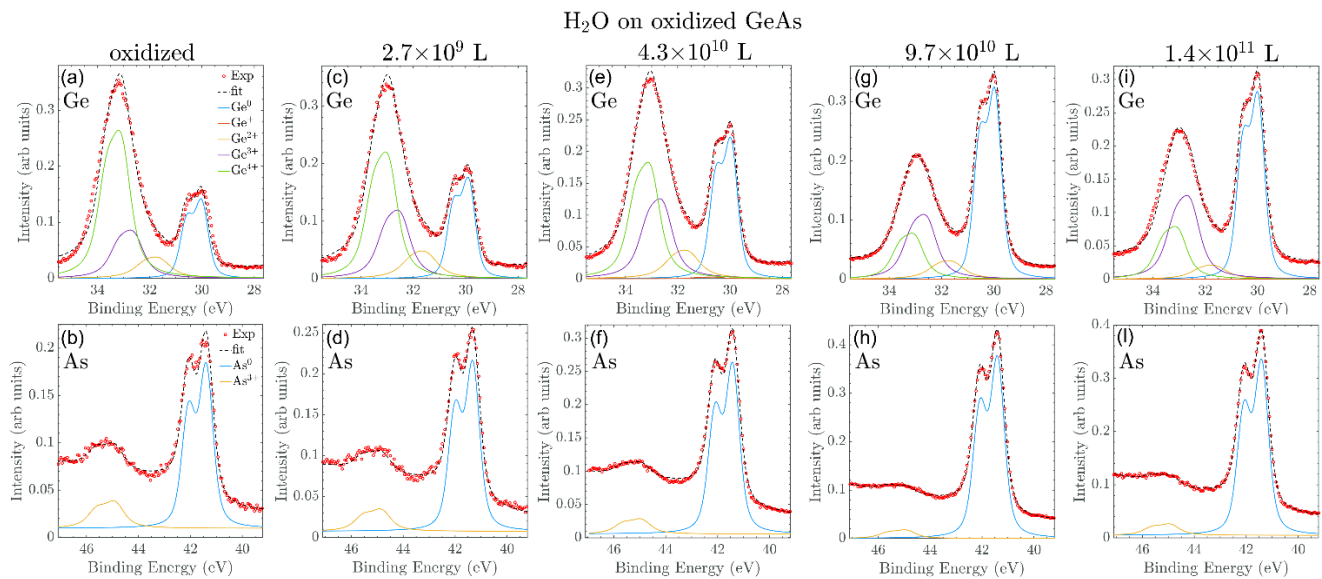

**Fig. 6. XPS spectra of (top row) Ge 3d and (bottom row) As 3d core levels of oxidized GeAs after exposure to $H_2O$ at increasing dosage: (a, b) initially oxidized sample; (c, d) $2.7\times10^9$ L; (e, f) $4.3\times10^{10}$ L; (g, h) $9.7\times10^{10}$ L; (i, l) $1.4\times10^{11}$ L. Spectra are measured with $E_p$= 700 eV and normalized to the total Ge 3d area at each dosage.**

To further confirm this effect, we carried out an experiment where an oxidized GeAs sample was exposed to $H_2O$ vapor at 15 Torr in dark conditions. Before such experiment, the sample was heavily oxidized ($L$= 1260 pm) by a prolonged dosage (about $2.0\times10^{10}$ L) of $O_2$ under the beam. Then, three consecutive exposures to $H_2O$ vapor were performed in dark conditions, measuring XPS after each of them (Fig. 6). Given the thick oxide layer, we used a photon energy $E_p$= 700 eV, resulting in an escape depth of about 2.8 nm. In the initial condition [Figs. 6(a) and 6(b)], the dominant germanium oxide component was $Ge^{4+}$, whilst for arsenic oxide $As^{3+}$ was the only one present. The appearance of high oxidation numbers for germanium and arsenic is in line with the heavy oxidation condition. Monitoring the evolution during exposure to water vapor reveals that only $Ge^{4+}$ is reduced as the dosage increases, while the $GeO_x$ suboxide components ($Ge^{3+}$, $Ge^{2+}$) remain nearly constant [Fig. 7(a)]. This indicates that the reaction of water does not reduce $Ge^{4+}$ oxide species to a lower oxidation number but completely and selectively removes $GeO_2$. Consequently, the relative amount of unoxidized $Ge^0$ is increased, and the equivalent germanium oxide thickness almost halved, decreasing to about 740 pm at $1.4\times10^{10}$ L. As for arsenic, the relative area intensity of the $As^{3+}$ oxide component is slightly diminished as the $H_2O$ vapor dosage is increased [Fig. 7(b)]. It is however more striking to note the concomitant increase of the relative total arsenic content (with respect to germanium) in the sample, as highlighted in Fig. 7(c). The initial oxidized sample is in fact poor in arsenic with respect to the GeAs pristine stoichiometry, due to the loss of volatile AsOx which, as seen earlier, occurs at the early stages of oxidation. This change in stoichiometry, however, is in turn partially restored by the loss of water-soluble $GeO_2$ occurring during exposure to water.

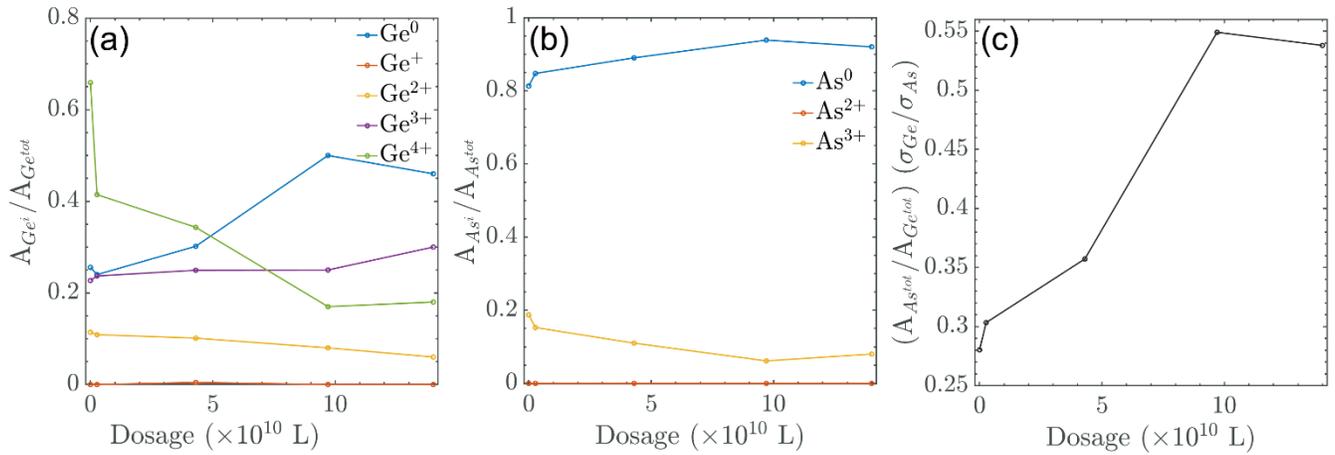

**Fig. 7. Oxidized GeAs sample exposed to $H_2O$. Evolution with dosage of: (a) area of each Ge 3d fitting component with respect to the total Ge 3d area; (b) area of each As 3d fitting component with respect to the total As 3d area; (c) stoichiometry obtained from the ratio of the areas of the As 3d and Ge 3d peaks normalized by the corresponding photoionization cross sections. $E_p$= 700 eV.**

*Discussion*

In the previous section, we observed that a common characteristic of GeAs oxidation, upon exposure to oxygen, water, or both, is the depletion of arsenic, which consistently accompanies the formation of germanium oxide. This correlation is made evident in Fig. 8(a). The plot is obtained from all the data measured with the most-surface sensitive photon energy of 350 eV. The horizontal axis corresponds to the $A_{Ge^0}/A_{Ge^{tot}}$ ratio and, thus, ranges between 1 (unoxidized germanium of pristine GeAs) and 0 (all the germanium in GeAs being completely oxidized). The vertical axis represents the stoichiometric ratio of arsenic to germanium, with a value of 1 for pristine GeAs, whereas lower values indicate arsenic depletion in the material. It is clear that as more arsenic is lost, the oxidation of germanium increases. In other words, oxidation of GeAs proceeds through loss of As and consequent oxidation of Ge, at least at the initial stages of the reaction. We can also notice, as already mentioned, that the arsenic loss is typically less than 20 % which means, from the electron escape depth, that only the topmost arsenic atoms in the first layer of GeAs are involved in the reaction.

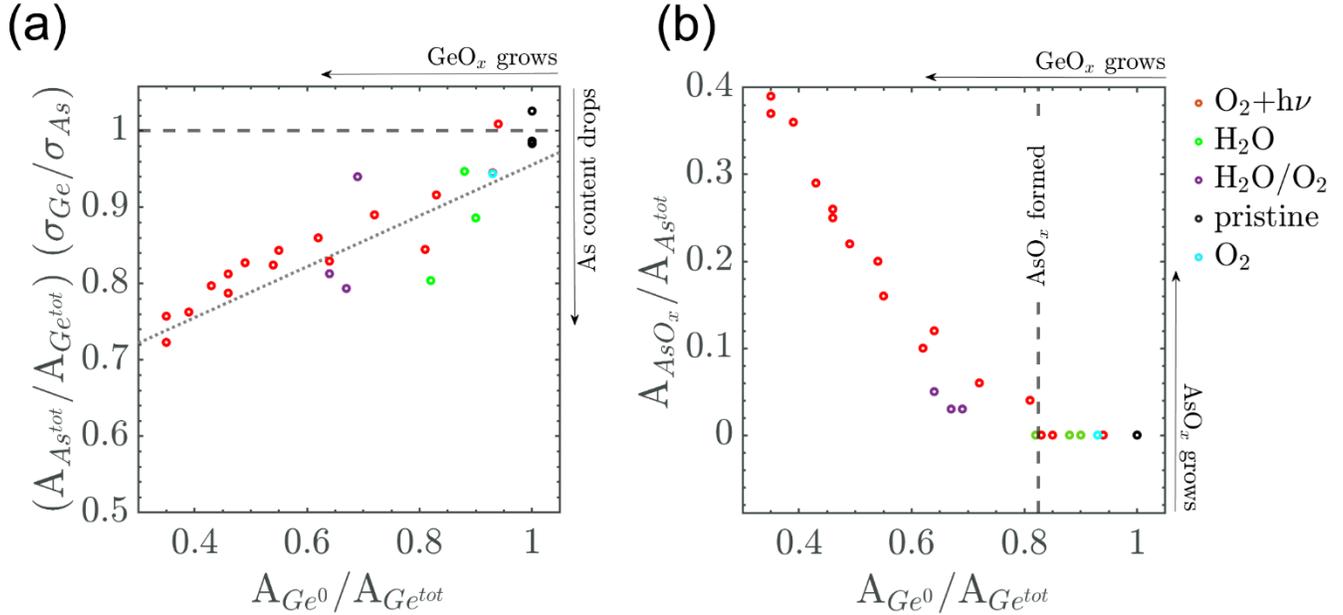

**Fig. 8. (a) Correlation between the stoichiometry obtained from the ratio of the areas of the As 3d and Ge 3d peaks (normalized by the corresponding photoionization cross sections) and the relative area intensity of the unoxidized Ge$^0$ component. (b) Correlation between the relative area intensity of the arsenic oxide components (considering both As$^{2+}$ and As$^{3+}$) and the relative area intensity of the unoxidized Ge$^0$ component. Data obtained with a photon energy of 350 eV.**

Spectroscopic data obtained with $E_p$= 350 eV were further analyzed by plotting the relative area intensity of the arsenic oxide components (considering both As$^{2+}$ and As$^{3+}$) versus the normalized area of the unoxidized Ge$^0$ component [Fig. 8(b)]. It is observed that the initial decrease in the Ge$^0$ component (i.e. the formation of GeO$_x$ species) is not immediately followed by the formation of arsenic oxide. Instead, arsenic oxide begins to form only when the attenuation of the Ge$^0$ reaches approximately 20%. This, as previously mentioned, equates to a germanium oxide thickness roughly equal to 1/5 of the top GeAs layer, corresponding to the oxidation of the uppermost germanium atoms in that layer. Only after a sufficiently thick germanium oxide layer is formed, arsenic oxide begins to be observed. In the case of the exposure to $H_2O$ vapor or $O_2$ in dark conditions, the germanium oxide layer thickness is below this threshold and no arsenic oxide is therefore detected.

To support our experimental findings, we performed theoretical studies on the dissociative chemisorption of $O_2$, $H_2O$, and their combined presence on GeAs. The results are displayed in Fig. 9. In agreement with the experiment, we find that, independent of the reacting gas, the attack on GeAs indeed starts at the topmost As atoms of the first GeAs layer. As shown in Fig. 9(a), the chemisorption of $O_2$ results in being only slightly exothermic ($E_f$= -0.2 eV, following eq. (1), X=$O_2$). Considering the energy accuracy of DFT-PBA for reactions involving molecular species[53], it is reasonable to conclude that the oxygen-initiated oxidation of GeAs is at the threshold of thermodynamic stability, consistent with the experimental results. On the other end, the chemisorption of one $H_2O$ molecule is a quite endothermic process with the most energetically favorable attack (H—Ge ••• As—OH), in Fig. 9(b), showing $E_f$= +0.88 eV. We also investigated another attack (H—As ••• Ge—OH, see Fig. S5), obtaining an even larger formation energy for the chemisorbed species of +1.05 eV. As discussed in the previous section, our experiment involved exposing the sample to water vapor at a pressure high enough to form a liquid water film on the surface of GeAs. Under these conditions, the presence of physisorbed $H_2O$ molecules surrounding the chemisorbed species cannot be ignored. To account for this effect, we conducted a simulation where one $H_2O$ molecule was chemisorbed while additional (up to three) physisorbed $H_2O$ molecules were situated nearby [Figs. 9(c) to 9(e)]. Interestingly, we discovered that the formation energy for the process significantly decreases as the number of physisorbed $H_2O$ molecules increases. In fact, it becomes negative (exothermic) when two physisorbed molecules are near the chemisorbed one, and even more exothermic with three physisorbed molecules. Although the actual experimental conditions are reasonably more complex, theoretical insights thus suggest that the presence of physisorbed water molecules on the GeAs surface, occurring under realistic environmental conditions, plays a crucial role in reducing the energy required for water chemisorption.

In the previous section, we found that the reactivity of GeAs was significantly higher when exposed simultaneously to both $O_2$ and $H_2O$ compared to either reactant gas alone. Given the 12:1 ratio of $O_2$ to $H_2O$ used in the experiment, we simulated the chemical adsorption of $H_2O$ following the oxidation of GeAs initiated by $O_2$ [Fig. 9(f)]. In a perfect match with the experiment, this simulation yielded a highly exothermic process ($E_f$= -2.18 eV), which is more exothermic than any other interaction discussed thus far. Interestingly, we also observed that the Ge—As bond distance increases when water is chemically absorbed (the longest Ge—As being 2.81 Å) compared with the same bond distance in the initially $O_2$ oxidized system (Ge—As: 2.77 Å). The increase in the bond length could facilitate the loss of arsenic atoms which was indeed observed in the experiment.

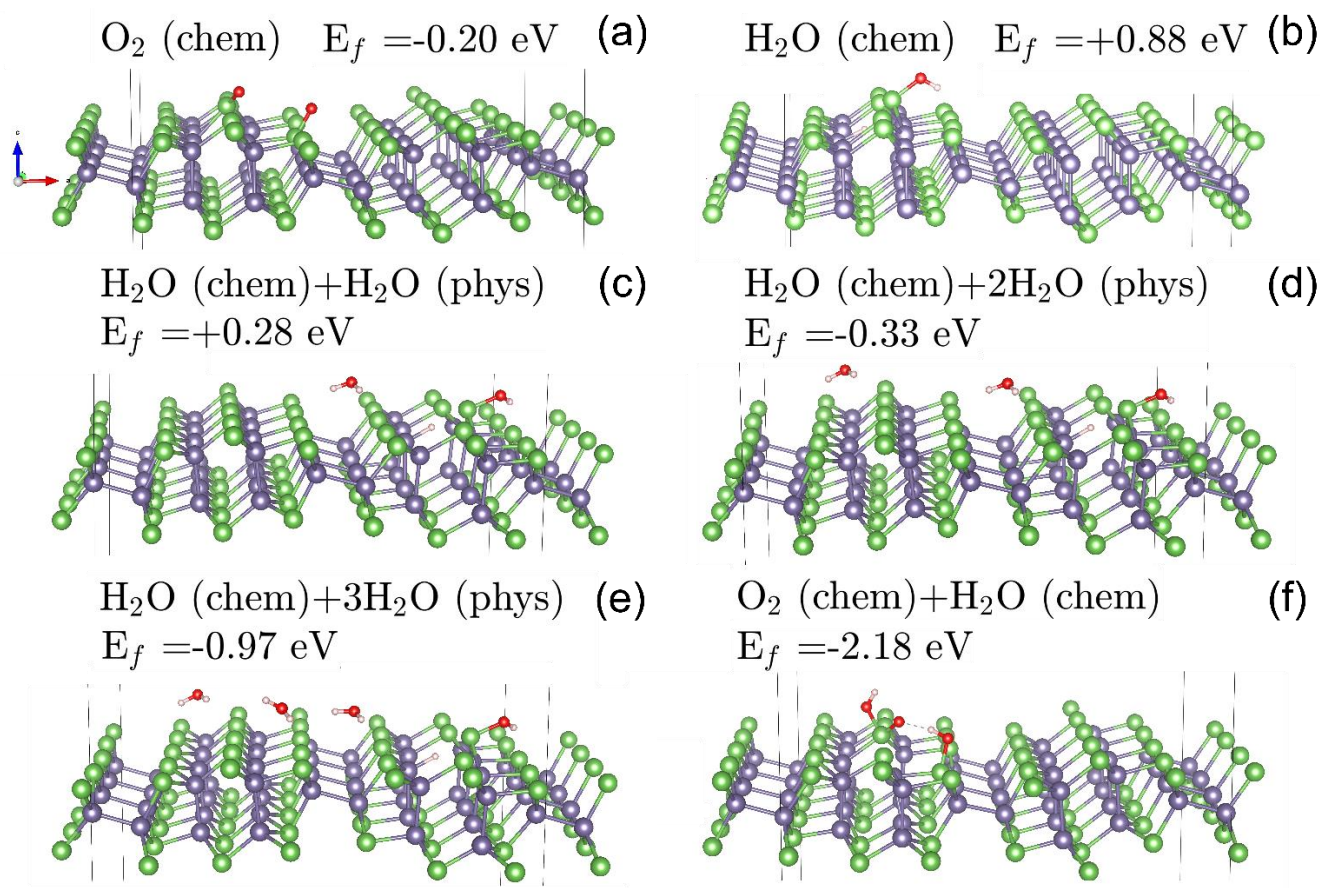

**Fig. 9. PAW/PBE optimized structures of the adducts formed by GeAs with $O_2$ and $H_2O$ along with the corresponding formation energies . (a) Chemisorption of one $O_2$ molecule; (b) chemisorption of one $H_2O$ molecule; (c) chemisorption of one $H_2O$ molecule in the presence of a second nearby physisorbed $H_2O$ molecule; (d) chemisorption of one $H_2O$ molecule in the presence of two nearby physisorbed $H_2O$ molecules; (e) chemisorption of one $H_2O$ molecule in the**

**presence of three nearby physisorbed $H_2O$ molecules; (f) concurrent chemisorption of $O_2$ and $H_2O$ molecules. $H_2O$ is chemisorbed on top of the $O_2$-oxidated structure in (a). [Green: As; Gray: Ge; Red: O; White: H atoms]**

## 5. Conclusions

We investigated the oxidation of 2D GeAs in realistic environmental conditions by combining synchrotron-based NAP XPS experiments with *ab initio* calculations. GeAs showed limited reactivity with dry $O_2$ and de-aerated $H_2O$, while the reactivity of the material is severely heightened by the presence of a small amount of humidity ($R_w$= 0.5% at *T*= 20 °C) in $O_2$ atmosphere. This critical interplay between oxygen and water on the surface of GeAs was accurately reproduced by DFT calculations which showed a strong exothermic formation energy for the concurrent chemisorption of both $O_2$ and $H_2O$, in contrast to that of the individual reactants. This finding is extremely important as any uncapped device working in real conditions will be subjected to the simultaneous attack of both oxygen and water.

XPS data also revealed that the $Ge^{4+}$ oxide species, which correspond to Ge dioxide, are selectively removed upon exposure to water. This offers support to a technological application of water-etching of germanium-based 2D semiconductors, including GeAs, GeS, and $GeS_2$, that has been recently proposed to control the thickness of these materials[45]. This underscores the significance of our findings in understanding the surface chemistry of these materials when interacting with water. These interactions are crucial not only to improve the longevity and stability of GeAs-based devices, but also for the development of scalable liquid-phase exfoliation and dispersion methods.

**Acknowledgements**

We acknowledge SOLEIL for provision of synchrotron radiation facilities for the beamtime 20230131 and would like to thank the TEMPO beamline's staff for assistance. We acknowledge funding support from the Italian Ministry of Research through the project Quantum Transition-metal FLUOrides (QT-FLUO) PRIN 20207ZXT4Z and the projects P2022LXNYN and 2022JW8LHZ within the European

Union-Next Generation EU program. G.G. acknowledges the CINECA award under the ISCRA initiative, for the availability of high-performance computing resources and support. G.G. acknowledges financial support under the National Recovery and Resilience Plan (NRRP), Mission 4, Component 2, Investment 1.1, Call for tender No. 104 published on 2.2.2022 by the Italian Ministry of University and Research (MUR), funded by the European Union – NextGenerationEU – Project Title 2022LZWKAJ Engineered nanoheterostructures for a new generation of titania photocatalytic films (ENTI) – CUP J53D23003470006 - Grant Assignment Decree No. 966 adopted on 30.06.2023 by the Italian Ministry of University and Research (MUR). G.G. also thanks the European Union - NextGenerationEU under the Italian Ministry of University and Research (MUR) National Innovation Ecosystem Grant ECS00000041 - VITALITY for funding and acknowledges Università degli Studi di Perugia and MUR for support within the project Vitality.

**References**

(**1**) Lin, Y.-C.; Torsi, R.; Younas, R.; Hinkle, C. L.; Rigosi, A. F.; Hill, H. M.; Zhang, K.; Huang, S.; Shuck, C. E.; Chen, C.; et al. Recent Advances in 2D Material Theory, Synthesis, Properties, and Applications. *ACS Nano* **2023**, *17* (11), 9694-9747. DOI: 10.1021/acsnano.2c12759.

(**2**) Katiyar, A. K.; Hoang, A. T.; Xu, D.; Hong, J.; Kim, B. J.; Ji, S.; Ahn, J. H. 2D Materials in Flexible Electronics: Recent Advances and Future Prospectives. *Chem Rev* **2024**, *124* (2), 318-419. DOI: 10.1021/acs.chemrev.3c00302  From NLM.

(**3**) Liu, A.; Zhang, X.; Liu, Z.; Li, Y.; Peng, X.; Li, X.; Qin, Y.; Hu, C.; Qiu, Y.; Jiang, H.; et al. The Roadmap of 2D Materials and Devices Toward Chips. *Nano-Micro Letters* **2024**, *16* (1), 119. DOI: 10.1007/s40820-023-01273-5.

(**4**) Cho, Y. S.; Kang, J. Two-dimensional materials as catalysts, interfaces, and electrodes for an efficient hydrogen evolution reaction. *Nanoscale* **2024**, *16* (8), 3936-3950, 10.1039/D4NR00147H. DOI: 10.1039/D4NR00147H.

(**5**) Abbas, A.; Luo, Y.; Ahmad, W.; Mustaqeem, M.; Kong, L.; Chen, J.; Zhou, G.; Tabish, T. A.; Zhang, Q.; Liang, Q. Recent progress, challenges, and opportunities in 2D materials for flexible displays. *Nano Today* **2024**, *56*, 102256. DOI: https://doi.org/10.1016/j.nantod.2024.102256.

(**6**) Xue, Y.; Xu, T.; Wang, C.; Fu, L. Recent advances of two-dimensional materials-based heterostructures for rechargeable batteries. *iScience* **2024**, *27* (8), 110392. DOI: https://doi.org/10.1016/j.isci.2024.110392.

**(7)** Hu, Z.; Li, Q.; Lei, B.; Zhou, Q.; Xiang, D.; Lyu, Z.; Hu, F.; Wang, J.; Ren, Y.; Guo, R.; et al. Water-Catalyzed Oxidation of Few-Layer Black Phosphorous in a Dark Environment. *Angewandte Chemie International Edition* **2017**, *56* (31), 9131-9135. DOI: https://doi.org/10.1002/anie.201705012.

**(8)** Huang, Y.; Qiao, J.; He, K.; Bliznakov, S.; Sutter, E.; Chen, X.; Luo, D.; Meng, F.; Su, D.; Decker, J.; et al. Interaction of Black Phosphorus with Oxygen and Water. *Chem. Mater.* **2016**, *28* (22), 8330-8339. DOI: 10.1021/acs.chemmater.6b03592.

**(9)** Zhou, Q.; Chen, Q.; Tong, Y.; Wang, J. Light-Induced Ambient Degradation of Few-Layer Black Phosphorus: Mechanism and Protection. *Angewandte Chemie International Edition* **2016**, *55* (38), 11437-11441. DOI: https://doi.org/10.1002/anie.201605168.

**(10)** Wang, X.; Sun, Y.; Liu, K. Chemical and structural stability of 2D layered materials. *2D Materials* **2019**, *6* (4), 042001. DOI: 10.1088/2053-1583/ab20d6.

**(11)** Tanwar, M.; Udyavara, S.; Yun, H.; Ghosh, S.; Mkhoyan, K. A.; Neurock, M. Co-Operative Influence of $O_2$ and $H_2O$ in the Degradation of Layered Black Arsenic. *The Journal of Physical Chemistry C* **2022**, *126* (36), 15222-15228. DOI: 10.1021/acs.jpcc.2c04861.

**(12)** Bussolotti, F.; Kawai, H.; Maddumapatabandi, T. D.; Fu, W.; Khoo, K. H.; Goh, K. E. J. Role of S-Vacancy Concentration in Air Oxidation of $WS_2$ Single Crystals. *ACS Nano* **2024**, *18* (12), 8706-8717. DOI: 10.1021/acsnano.3c10389.

**(13)** Dong, C.; Lu, L.-S.; Lin, Y.-C.; Robinson, J. A. Air-Stable, Large-Area 2D Metals and Semiconductors. *ACS Nanoscience Au* **2024**, *4* (2), 115-127. DOI: 10.1021/acsnanoscienceau.3c00047.

**(14)** Zhou, L.; Guo, Y.; Zhao, J. GeAs and SiAs monolayers: Novel 2D semiconductors with suitable band structures. *Physica E: Low-dimensional Systems and Nanostructures* **2018**, *95*, 149-153. DOI: https://doi.org/10.1016/j.physe.2017.08.016.

**(15)** Mortazavi, B.; Shahrokhi, M.; Cuniberti, G.; Zhuang, X. Two-Dimensional SiP, SiAs, GeP and GeAs as Promising Candidates for Photocatalytic Applications. *Coatings* **2019**, *9* (8), 522.

**(16)** Guo, J.; Liu, Y.; Ma, Y.; Zhu, E.; Lee, S.; Lu, Z.; Zhao, Z.; Xu, C.; Lee, S.-J.; Wu, H.; et al. Few-Layer GeAs Field-Effect Transistors and Infrared Photodetectors. *Advanced Materials* **2018**, *30* (21), 1705934. DOI: https://doi.org/10.1002/adma.201705934.

**(17)** Jiang, X.; Zhao, T.; Wang, D. Anisotropic ductility and thermoelectricity of van der Waals GeAs. *Phys. Chem. Chem. Phys.* **2023**, *25* (40), 27542-27552, 10.1039/D3CP03119E. DOI: 10.1039/D3CP03119E.

**(18)** Barreteau, C.; Michon, B.; Besnard, C.; Giannini, E. High-pressure melt growth and transport properties of SiP, SiAs, GeP, and GeAs 2D layered semiconductors. *J. Cryst. Growth* **2016**, *443*, 75-80. DOI: https://doi.org/10.1016/j.jcrysgro.2016.03.019.

**(19)** Yang, S.; Yang, Y.; Wu, M.; Hu, C.; Shen, W.; Gong, Y.; Huang, L.; Jiang, C.; Zhang, Y.; Ajayan, P. M. Highly In-Plane Optical and Electrical Anisotropy of 2D Germanium Arsenide. *Adv. Funct. Mater.* **2018**, *28* (16), 1707379. DOI: https://doi.org/10.1002/adfm.201707379.

**(20)** Lee, K.; Kamali, S.; Ericsson, T.; Bellard, M.; Kovnir, K. GeAs: Highly Anisotropic van der Waals Thermoelectric Material. *Chem. Mater.* **2016**, *28* (8), 2776-2785. DOI: 10.1021/acs.chemmater.6b00567.

**(21)** Hoat, D. M.; Ponce-Pérez, R.; Ha, C. V.; Guerrero-Sanchez, J. Controlling the electronic and magnetic properties of the GeAs monolayer by generating Ge vacancies and doping with transition-metal atoms. *Nanoscale Advances* **2024**, *6* (14), 3602-3611. DOI: https://doi.org/10.1039/d4na00235k.

**(22)** Zhou, Z.; Long, M.; Pan, L.; Wang, X.; Zhong, M.; Blei, M.; Wang, J.; Fang, J.; Tongay, S.; Hu, W.; et al. Perpendicular Optical Reversal of the Linear Dichroism and Polarized Photodetection in 2D GeAs. *ACS Nano* **2018**, *12* (12), 12416-12423. DOI: 10.1021/acsnano.8b06629.

**(23)** Dushaq, G.; Villegas, J. E.; Paredes, B.; Tamalampudi, S. R.; Rasras, M. S. Anisotropic Van Der Waals 2D GeAs Integrated on Silicon Four-Waveguide Crossing. *J. Lightwave Technol.* **2023**, *41* (6), 1784-1789. DOI: 10.1109/JLT.2022.3229069.

**(24)** Sar, H.; Gao, J.; Yang, X. In-plane anisotropic third-harmonic generation from germanium arsenide thin flakes. *Scientific Reports* **2020**, *10* (1), 14282. DOI: 10.1038/s41598-020-71244-y.

**(25)** Sun, J.; Passacantando, M.; Palummo, M.; Nardone, M.; Kaasbjerg, K.; Grillo, A.; Di Bartolomeo, A.; Caridad, J. M.; Camilli, L. Impact of Impurities on the Electrical Conduction of Anisotropic Two-Dimensional Materials. *Physical Review Applied* **2020**, *13* (4), 044063. DOI: 10.1103/PhysRevApplied.13.044063.

**(26)** Kim, J. H.; Moon, B. H.; Han, G. H. Anti-ambipolar transport and logic operation in two-dimensional field-effect transistors using in-series integration of GeAs and SnS$_2$. *Appl. Phys. Lett.* **2024**, *124* (12). DOI: 10.1063/5.0197983 (acccessed 9/18/2024).

**(27)** Kim, J. H.; Han, G. H.; Moon, B. H. GeAs as an emerging p-type van der Waals semiconductor and its application in p-n photodiodes. *Nanotechnol.* **2023**, *34* (31). DOI: 10.1088/1361-6528/acd1f5 From NLM.

**(28)** Grillo, A.; Di Bartolomeo, A.; Urban, F.; Passacantando, M.; Caridad, J. M.; Sun, J.; Camilli, L. Observation of 2D Conduction in Ultrathin Germanium Arsenide Field-Effect Transistors. *ACS Applied Materials & Interfaces* **2020**, *12* (11), 12998-13004. DOI: 10.1021/acsami.0c00348.

**(29)** Xiong, J.; Dan, Z.; Li, H.; Li, S.; Sun, Y.; Gao, W.; Huo, N.; Li, J. Multifunctional GeAs/WS$_2$ Heterojunctions for Highly Polarization-Sensitive Photodetectors in the Short-Wave Infrared Range. *ACS Applied Materials & Interfaces* **2022**, *14* (19), 22607-22614. DOI: 10.1021/acsami.2c03246.

**(30)** Grillo, A.; Faella, E.; Giubileo, F.; Pelella, A.; Urban, F.; Di Bartolomeo, A. Temperature Dependence of Germanium Arsenide Field-Effect Transistors Electrical Properties. *Materials Proceedings* **2021**, *4* (1), 26.

**(31)** Song, W.; Liu, H.; Zou, F.; Niu, Y.; Zhao, Y.; Cong, Y.; Pan, Y.; Li, Q. Isotropic Contact Properties in Monolayer GeAs Field-Effect Transistors. *Molecules* **2023**, *28* (23). DOI: 10.3390/molecules28237806 From NLM.

**(32)** Zhang, J.; Shang, C.; Dai, X.; Zhang, Y.; Zhu, T.; Zhou, N.; Xu, H.; Yang, R.; Li, X. Effective Passivation of Anisotropic 2D GeAs via Graphene Encapsulation for Highly Stable Near-Infrared Photodetectors. *ACS Applied Materials & Interfaces* **2023**, *15* (10), 13281-13289. DOI: 10.1021/acsami.2c20030.

**(33)** CRC Handbook of Chemistry and Physics: A Ready-Reference of Chemical and Physical Data, 85th ed Edited by David R. Lide (National Institute of Standards and Technology). CRC Press LLC: Boca Raton, FL. 2004. 2712 pp. $139.99. ISBN 0-8493-0485-7. *J. Am. Chem. Soc.* **2005**, *127* (12), 4542-4542. DOI: 10.1021/ja041017a.

**(34)** Stevie, F. A.; Donley, C. L. Introduction to x-ray photoelectron spectroscopy. *Journal of Vacuum Science & Technology A* **2020**, *38* (6). DOI: 10.1116/6.0000412 (acccessed 10/3/2024).

**(35)** Camilli, L.; Galbiati, M.; Di Gaspare, L.; De Seta, M.; Píš, I.; Bondino, F.; Caporale, A.; Veigang-Radulescu, V. P.; Babenko, V.; Hofmann, S.; et al. Tracking interfacial changes of graphene/Ge(110) during in-vacuum annealing. *Appl. Surf. Sci.* **2022**, *602*, 154291-154291. DOI: https://doi.org/10.1016/j.apsusc.2022.154291.

**(36)** Moreno, M.; Kumar, A.; Tallarida, M.; Horn, K.; Ney, A.; Ploog, K. H. Electronic states in arsenic-decapped MnAs (1-100) films grown on GaAs(001): A photoemission spectroscopy study. *Appl. Phys. Lett.* **2008**, *92* (8). DOI: 10.1063/1.2888953 (acccessed 9/18/2024).

**(37)** Kresse, G.; Hafner, J. Ab initio molecular dynamics for open-shell transition metals. *Phys. Rev. B* **1993**, *48* (17), 13115-13118. DOI: 10.1103/PhysRevB.48.13115.

**(38)** Kresse, G.; Hafner, J. Ab initio molecular-dynamics simulation of the liquid-metal--amorphous-semiconductor transition in germanium. *Phys. Rev. B* **1994**, *49* (20), 14251-14269. DOI: 10.1103/PhysRevB.49.14251.

**(39)** Kresse, G.; Furthmüller, J. Efficiency of ab-initio total energy calculations for metals and semiconductors using a plane-wave basis set. *Comp. Mater. Sci.* **1996**, *6* (1), 15-50. DOI: http://dx.doi.org/10.1016/0927-0256(96)00008-0.

**(40)** Kresse, G.; Furthmüller, J. Efficient iterative schemes for *ab initio* total-energy calculations using a plane-wave basis set. *Phys. Rev. B* **1996**, *54* (16), 11169-11186.

**(41)** Perdew, J. P.; Burke, K.; Ernzerhof, M. Generalized gradient approximation made simple. *Phys. Rev. Lett.* **1996**, *77*. DOI: 10.1103/PhysRevLett.77.3865.

**(42)** Blöchl, P. E. Projector augmented-wave method. *Phys. Rev. B* **1994**, *50* (24), 17953-17979.

**(43)** Grimme, S.; Antony, J.; Ehrlich, S.; Krieg, H. A consistent and accurate ab initio parametrization of density functional dispersion correction (DFT-D) for the 94 elements H-Pu. *The Journal of Chemical Physics* **2010**, *132* (15). DOI: 10.1063/1.3382344 (acccessed 9/18/2024).

**(44)** Camilli, L.; Capista, D.; Tomellini, M.; Sun, J.; Zeller, P.; Amati, M.; Gregoratti, L.; Lozzi, L.; Passacantando, M. Formation of a two-dimensional oxide via oxidation of a layered material. *Phys. Chem. Chem. Phys.* **2022**, *24* (22), 13935-13940, 10.1039/D2CP00863G. DOI: 10.1039/D2CP00863G.

**(45)** Sun, J.; Giorgi, G.; Palummo, M.; Sutter, P.; Passacantando, M.; Camilli, L. A Scalable Method for Thickness and Lateral Engineering of 2D Materials. *ACS Nano* **2020**, *14* (4), 4861-4870. DOI: 10.1021/acsnano.0c00836.

**(46)** Molle, A.; Bhuiyan, M. N. K.; Tallarida, G.; Fanciulli, M. In situ chemical and structural investigations of the oxidation of Ge(001) substrates by atomic oxygen. *Appl. Phys. Lett.* **2006**, *89* (8). DOI: 10.1063/1.2337543 (acccessed 9/18/2024).

**(47)** Wang, X.; Zhao, Z.; Xiang, J.; Wang, W.; Zhang, J.; Zhao, C.; Ye, T. Experimental investigation on oxidation kinetics of germanium by ozone. *Appl. Surf. Sci.* **2016**, *390*, 472-480. DOI: https://doi.org/10.1016/j.apsusc.2016.08.123.

**(48)** Budz, H. A.; Biesinger, M. C.; LaPierre, R. R. Passivation of GaAs by octadecanethiol self-assembled monolayers deposited from liquid and vapor phases. *Journal of Vacuum Science & Technology B: Microelectronics and Nanometer Structures Processing, Measurement, and Phenomena* **2009**, *27* (2), 637-648. DOI: 10.1116/1.3100266 (acccessed 9/18/2024).

**(49)** Rochet, F.; Poncey, C.; Dufour, G.; Roulet, H.; Rodrigues, W. N.; Sauvage, M.; Boulliard, J. C.; Sirotti, F.; Panaccione, G. The As-terminated Si(001) surface and its oxidation in molecular oxygen: an Si 2p and As 3d core-level study with synchrotron radiation. *Surf. Sci.* **1995**, *326* (3), 229-242. DOI: https://doi.org/10.1016/0039-6028(94)00793-4.

**(50)** NIST X-ray Photoelectron Spectroscopy Database, NIST Standard Reference Database Number 20, National Institute of Standards and Technology, Gaithersburg MD, 20899 (2000), DOI: https://dx.doi.org/10.18434/T4T88K.

**(51)** Abrenica, G. H. A.; Lebedev, M. V.; Fingerle, M.; Arnauts, S.; Bazzazian, N.; Calvet, W.; Porret, C.; Bender, H.; Mayer, T.; de Gendt, S.; et al. Atomic-scale investigations on the wet etching kinetics of Ge versus SiGe in acidic $H_2O_2$ solutions: a post operando synchrotron XPS analysis. *Journal of Materials Chemistry C* **2020**, *8* (29), 10060-10070, 10.1039/D0TC02763D. DOI: 10.1039/D0TC02763D.

**(52)** Grossi, V.; Ottaviano, L.; Santucci, S.; Passacantando, M. XPS and SEM studies of oxide reduction of germanium nanowires. *J. Non-Cryst. Solids* **2010**, *356* (37), 1988-1993. DOI: https://doi.org/10.1016/j.jnoncrysol.2010.05.042.

**(53)** Grindy, S.; Meredig, B.; Kirklin, S.; Saal, J. E.; Wolverton, C. Approaching chemical accuracy with density functional calculations: Diatomic energy corrections. *Phys. Rev. B* **2013**, *87* (7), 075150. DOI: 10.1103/PhysRevB.87.075150.

# Supporting Information

# Synergistic effect of oxygen and water on the environmental reactivity of 2D layered GeAs

*Luca Persichetti[1,±], Giacomo Giorgi[2,3,4,5], Luca Lozzi[6], Maurizio Passacantando[6,7], Fabrice Bournel[8,9], Jean-Jacques Gallet[8,9], Luca Camilli[1,*]*

[1]*Dipartimento di Fisica, Università di Roma "Tor Vergata", Via Della Ricerca Scientifica, 1- 00133 Rome, Italy*

[2]*Department of Civil and Environmental Engineering (DICA), University of Perugia, Via G. Duranti 93, 06125, Perugia, Italy;*

[3]*CIRIAF – Interuniversity Research Centre University of Perugia Via G. Duranti 93, 06125 Perugia, Italy*

[4]*CNR-SCITEC, 06123 Perugia, Italy*

[5]*Centro S3, CNR-Istituto Nanoscienze, Via G. Campi 213/a, Modena, 41125, Italy.*

[6]*Department of Physical and Chemical Science, University of L'Aquila, via Vetoio, Coppito, 67100 L'Aquila, Italy*

[7]*CNR-SPIN L'Aquila, via Vetoio, Coppito 67100, L'Aquila, Italy*

[8]*Sorbonne Université, CNRS, Laboratoire de Chimie Physique-Matière et Rayonnement, Campus Curie, UMR 7614, 4 place Jussieu, 75005 Paris, France*

[9]*Synchrotron SOLEIL, L'orme des Merisiers, B.P. 48, Saint Aubin, Gif-sur-Yvette Cedex 91192, France*

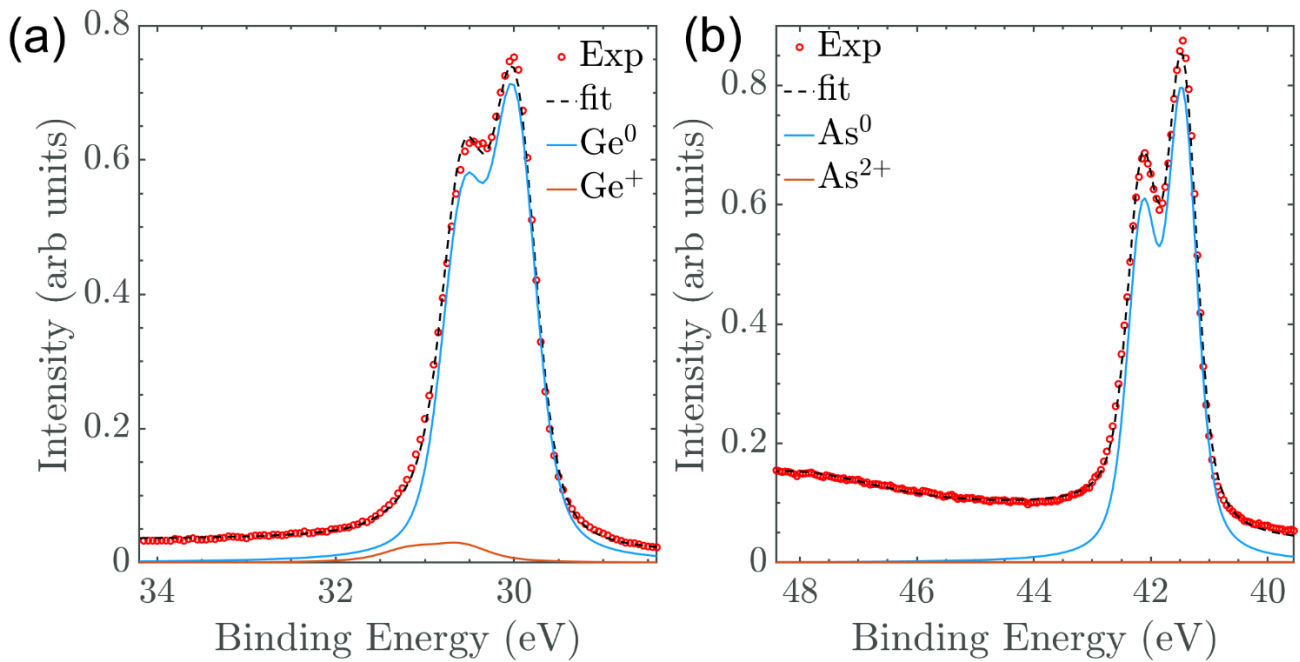

**Fig. S1.** XPS spectra of (a) Ge 3d and (b) As 3d core levels obtained with $E_p$= 500 eV after exposure of GeAs to $6.0\times10^{10}$ L of $O_2$.

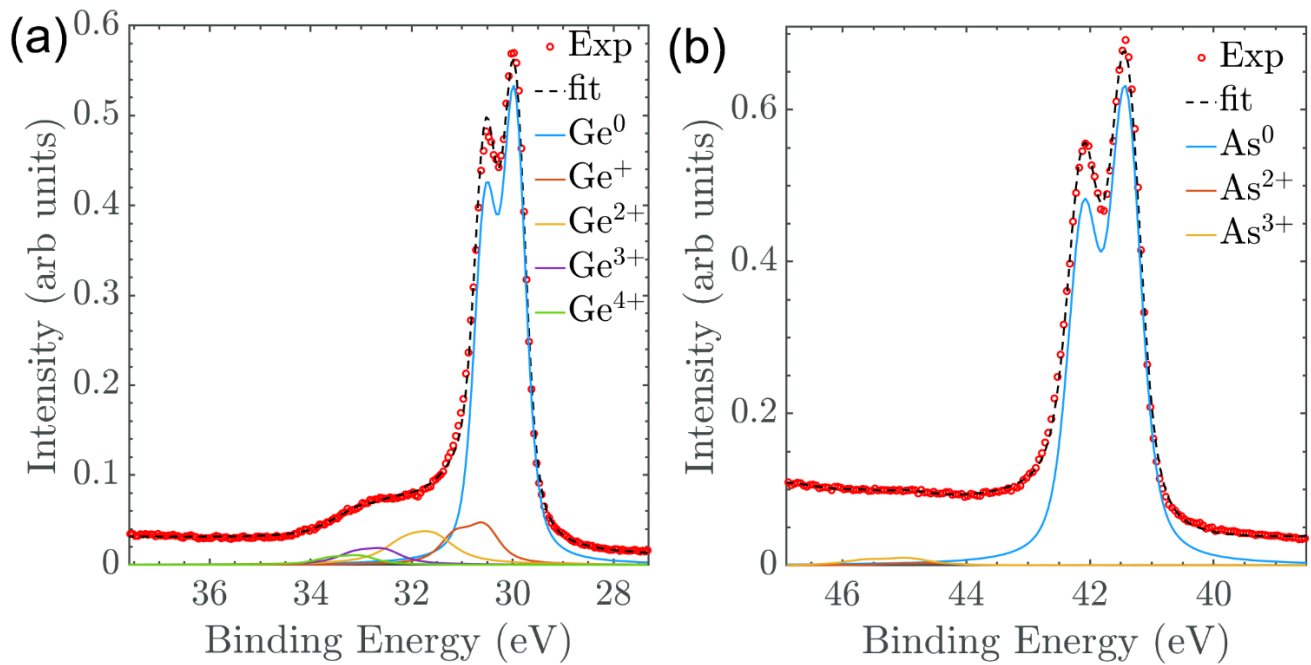

**Fig. S2.** XPS spectra of (a) Ge 3d and (b) As 3d core levels obtained with $E_p$= 500 eV after exposure of GeAs to $2.0\times10^{10}$ L of $O_2$+$H_2O$, $R_w$= 5% at 20 °C.

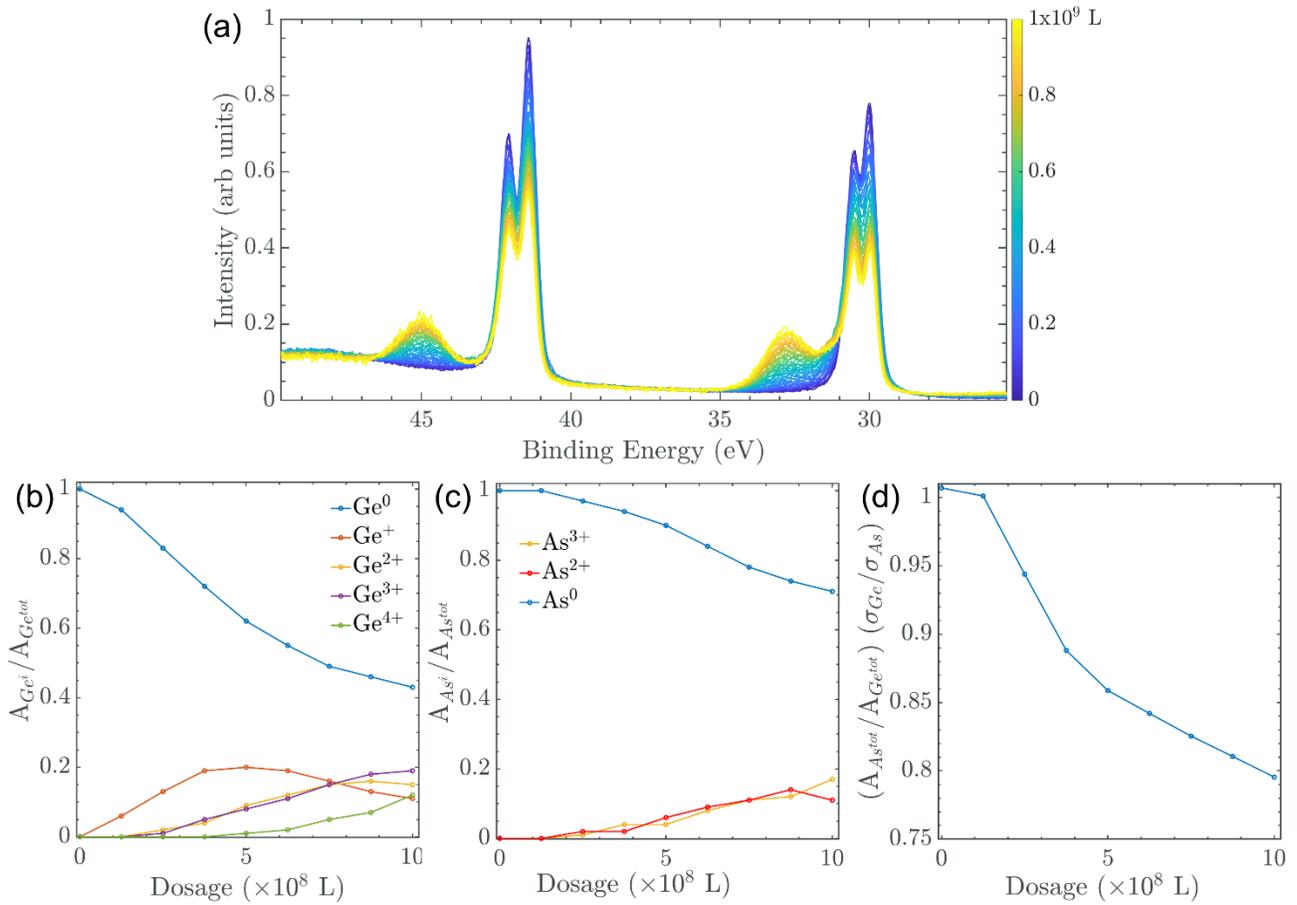

Fig. S3. (a) NAP-XPS spectra of Ge 3d and As 3d core levels measured with Ep= 350 eV while exposing GeAs to 0.15 Torr of $O_2$. (b-d) Evolution with dosage of: (b) area of each Ge 3d fitting component with respect to the total Ge 3d area; (c) area of each As 3d fitting component with respect to the total As 3d area; (d) stoichiometry obtained from the ratio of the areas of the As 3d and Ge 3d peaks normalized by the corresponding photoionization cross sections.

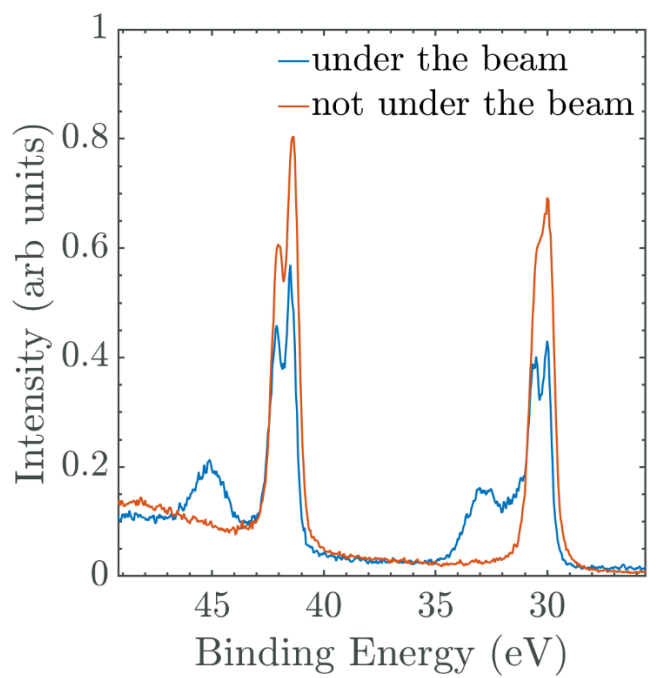

**Fig. S4.** XPS spectra of As 3d and Ge 3d core levels after dosing 4.0x10$^9$ L of O$_2$ taken in a sample location (blue) irradiated by the beam and in another one (red) not being irradiated by the beam during exposure.

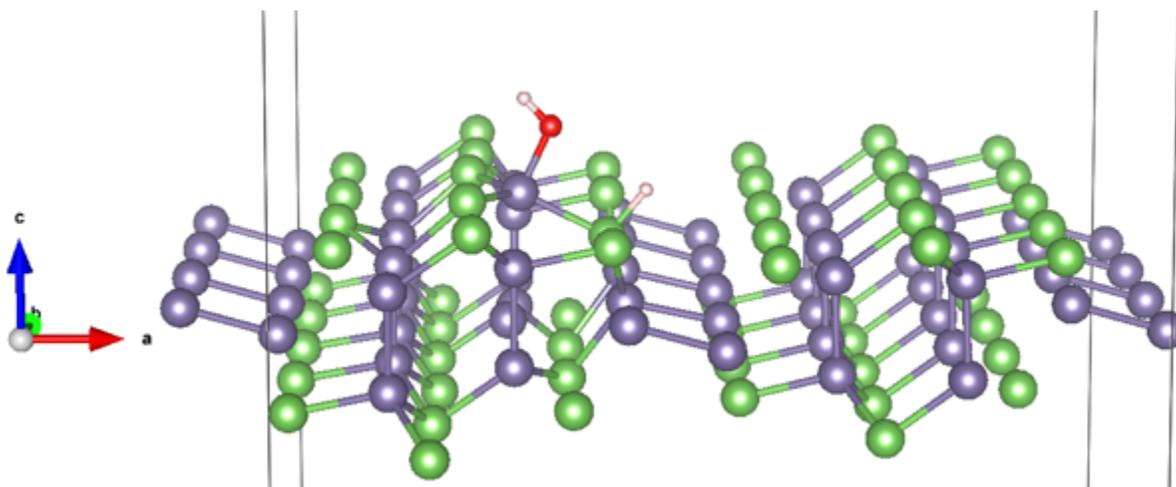

**Fig. S5.** PAW/PBE optimized structure of H$_2$O chemisorption on GeAs ML resulting in (H—As ••• Ge—OH). Here, E$_f$=+1.05 eV.

4